\documentclass[10pt, oneside]{article}   	
\usepackage{geometry}                		
\usepackage{graphicx}				
\usepackage{framed}				
\usepackage{amsmath,amssymb,mathtools}

\title{\bf Mean Time-to-Fire for the Noisy LIF Neuron \\ \medskip \sf \Large A detailed derivation of the Siegert Formula} 

\author{{Ken Kreutz-Delgado\textsuperscript{$\dagger$}}} 

\date{January 16, 2015} 

\begin{document}



\maketitle 

\setcounter{tocdepth}{2} 

%
%


\section*{Abstract} 
When stimulated by a very large number of Poisson-like presynaptic current input spikes, the temporal dynamics of the soma membrane potential $V(t)$ of a leaky integrate-and-fire (LIF) neuron is typically modeled in the diffusion limit and treated as a Ornstein-Uhlenbeck process (OUP).  When the potential reaches a threshold value $\theta$,  $V(t) = \theta$, the LIF neuron fires and the membrane potential is reset to a resting value, $V_0 < \theta$,  and clamped to this value for a specified (non-stochastic) absolute refractory period $T_r \ge 0$, after which the cycle is repeated.  The time between firings is given by the random variable $T_f = T_r+ T$ where $T$ is the random time which elapses between the ``unpinning'' of the membrane potential clamp and the next, subsequent firing of the neuron. The mean time-to-fire, $\widehat{T}_f = \text{E}(T_f) = T_r + \text{E}(T) = T_r + \widehat{T}$, provides a measure $\rho$ of the  average firing rate of the neuron,
\[ \rho = \widehat{T}_f^{-1} = \frac{1}{T_r + \widehat{T}} . \]
This note briefly discusses some aspects of the OUP model and derives the Siegert formula giving the firing rate, $\rho = \rho(I_0)$ as a function of an injected current, $I_0$.  This is a well-known classical result and no claim to originality is made. The derivation of the firing rate given in this report, which closely follows the derivation outlined by Gardiner \cite{cwGardiner:2009a}, minimizes the required mathematical background and is done in some pedagogic detail to facilitate study by graduate students and others who are new to the subject.  Knowledge of the material presented in the first five chapters of Gardiner \cite{cwGardiner:2009a} should provide an adequate background for following the derivation given in this note.


{\let\thefootnote\relax\footnotetext{\textsuperscript{$\dagger$} \textit{Department of Electrical \& Computer Engineering, University of California, San Diego. Email: kreutz@ece.ucsd.edu}}}





\section{Background}

Simple phenomenological models of the dynamical depolarization behavior of the membrane potential, $V(t)$, of a neuron often separate its behavior into two phases: (1) an analytically tractable ``subthreshold'' behavior, which describes the behavior until a threshold membrane potential value, $V(t) = \theta$, is attained. (2) At which point an action potential spike is triggered, followed by an absolute refractory period after which the cycle repeats itself.

\medskip
In particular, the \emph{leaky integrate and fire} (LIF) model assumes that the depolarization dynamics in the subthreshold regime of the part of the soma responsible for spike generation behaves like a linear $RC$ circuit,
\begin{align}
\boxed{\text{\bf LIF\, :} \qquad \frac{d}{dt}V(t)  = - \frac{1}{\tau} (V(t)  - V_r) + \frac{1}{C}\, I (t) \, , \qquad \tau = RC } \label{eq:LIF1}
\end{align}
This assumption is consistent with the fact that averaging over a length of a cylindrical neuron membrane yields this behavior as a low-order approximation \cite{hcTuckwell:1988v1,hcTuckwell:1989a}. For mammalian neurons a typical time-constant value is $\tau \sim$ 5--10 msec \cite{hcTuckwell:1988v1,dSterratt:2011a}.
 The constant membrane reversal (aka, equilibrium) potential is denoted by $V_r$.
 
\medskip
Notwithstanding the gross simplification that occurs from the use of the LIF,  additional significant approximation issues occur due to the nature of the stochastic assumptions that are made on the afferent currents which comprise the components of the membrane current $I$.  Generally we can write,
\[ I(t) = I_{\rm e}(t) + I_{\rm int} + I_{\rm syn} , \]
where $I_{\rm e}(t)$ is an externally injected current for experimental or control purposes, $I_{\rm int}$ represents internal or internal noise which is generally neglected,\footnote{Internal, intrinsic noise is thermal noise (Johnson or Nyquist noise), ``shot noise'' (due to ions clumping up as they pass through channels) and channel noise due to the stochastic natures of the opening and closing of large numbers of channel pores on the neuronal membrane. Even taken collectively, such noise affects are usually (but not always) of low order compared to the other terms.} while $I_{\rm syn}$ represents the summed effect of all of the afferent ``upstream'' neurons which are sending spikes to the neuron across synaptic clefts. Ignoring the intrinsic noise and treating the injected noise as constant on the time-scale of interest, $I_{\rm}(t) = I_0$, we have,
\begin{align}
I(t) = I_0 + I_{\rm syn} = I_0 + \sum_{j=1}^{N_{\rm syn}}  e_j I_j(t) ,  \label{eq:inputcurrents}
\end{align}
where $j$ indexes one of the $N_{\rm sun}$ synapses, $I_j(t)$ is a current induced at the $j$ synapse by presynaptic stimulation, and $e_j$ is the synaptic efficacy of the $j$ synapse which is positive for input from an excitatory presynaptic neuron and negative for an inhibitory presynaptic neuron.  As many researchers have noted \cite{rmCapocelli:1971a,nsGoel:1974a,lmRicciardi:1977a,hcTuckwell:1988v2,hcTuckwell:1989a,djAmit:1991a,wGerstner:2002a,aRenart:2003a} at the finest time scale one cannot ignore the fact that each $I_j(t)$ is effectively a spike train.\footnote{The resulting model is known as the \emph{Stein Model} \cite{rbStein:1965a,rbStein:1967a,hcTuckwell:1988v2,hcTuckwell:1989a}.}  However it is much more difficult to analyze a spiking stimulus than one that has continuous sample paths.  

\medskip
For this reason, as discussed in the aforementioned references, it is commonly argued that for neuronal ensembles that operate in the limit of high impinging spiking rates (relative to the membrane time constant) and small per-spike efficacies (relative to the magnitude of the spiking threshold magnitude) one can approximate a spiking point process which is assumed to have independent increments by a diffusive, continuous sample path Wiener process (aka Brownian motion).  And indeed, such arguments can at times be made rigorous  \cite{rmCapocelli:1971a,lmRicciardi:1977a,cwGardiner:2009a,vanKampen:2007a}.
The diffusion approximation is made by matching the mean and variance of a input spike train to the drift (local mean) and intensity (local variance) of a Wiener process.  In particular, if each of the input current spike trains represented by $I_j(t)$ is  a stationary poisson renewal process with rate $\lambda_j$, we have (see, e.g., \cite{hcTuckwell:1988v2}),
\begin{align}
  \widehat{I}_{\rm syn} = \text{E}\left\{ I_{\rm syn}(t) \right\} =  \sum_j e_j \lambda_j  \quad \text{and} \quad s^2_{\rm syn} = \text{Var}\left\{ I_{\rm syn}(t)  \right\}=
\sum_j e_j^2 \lambda_j   \label{eq:synmoments}
\end{align}
The practical reasonableness of approximating $I_{\rm syn}(t)$ by a Wiener process with this (local) mean and variance is discussed in \cite{djAmit:1991a,aRenart:2003a}.

\medskip
An additional problem arises because in order to model $I_{\rm syn}$ by a continuous sample path Wiener process, it should be an independent increment process.  However, there are capacitive dynamics at work in the synapses which induce correlations.  To rigorously model such effects is onerous, so an additional approximation is usually made.  In addition to assuming that the spike trains impinging on the soma are a  Wiener process, it is also assumed that the capacitive time constants at the synapses are fast compared to the time constant of the soma membrane. This assumption can also be problematic (see the discussions in \cite{aRenart:2003a,nBrunel:1998a}).  Suffice it to say, that in this note we do make the following huge simplifications for purposes of mathematical tractability:
\begin{itemize}
\item We assume that we can model the firing behavior of a neuron by the simple LIF model given by Eq.~\eqref{eq:LIF1}.
\item We assume that the collective effect of the  presynaptic spike trains impinging on the neuron can be model by an independent increment Wiener process.
\item We assume that the capacitive effects of synaptic and dendritic accumulation and processing of the spike trains are negligible. 
\end{itemize}

\medskip
To assemble the above pieces into the complete model which we will use in the sequel, define the zero mean process and unit variance process,
\[ \widetilde{I}_{\rm syn}(t) =  \frac{1}{s_{\rm syn}} \left( I (t) - \widehat{I}_{\rm syn} \right) , \]
so that,
\begin{align} I_{\rm syn}(t) = \widehat{I}_{\rm syn} + s_{\rm syn} \widetilde{I}_{\rm syn}(t) . \label{eq:whitecurrent}
\end{align}
Combinging Eq.s~\eqref{eq:LIF1}, \eqref{eq:inputcurrents} \eqref{eq:whitecurrent} we have,
\begin{align}
\frac{d}{dt}V(t)  = \underbrace{\underbrace{\frac{1}{\tau} V_r +  \frac{1}{C}\widehat{I}_{\rm syn}  + I_0}_{\alpha_0 = \mu}  - \underbrace{\frac{1}{\tau}}_{\alpha_1 = \kappa} V(t)}_{\alpha(V) = \alpha_0 + \alpha_1 V}   + \underbrace{\frac{s_{\rm syn}}{C}}_{\sqrt{\beta} = \sigma} \, \widetilde{I}_{\rm syn}(t).
\label{eq:alldefined}
\end{align}
Thus, 
\[ dV(t) = \alpha(V(t)) + \sqrt{\beta } \, d \widetilde{I}_{\rm syn}(t) . \]
To gain a further degree of abstraction, we set $X = V$ and $W = \widetilde{I}_{\rm syn}(t)$ where $W$ denotes the standard (zero mean and unit intensity) Wiener process.  More generally, $\beta$ can also depend on the state $X(t)$, $\beta = \beta(X)$.  The model,
\begin{align}
 dX(t) = \alpha(X(t)) + \sqrt{\beta(X(t)) \, \, } d W(t) , \label{eq:sdemodel}
\end{align}
is known as a stochastic differential equation \cite{cwGardiner:2009a,vanKampen:2007a} and has to be interpreted according to the rules of a stochastic calculus.\footnote{The standard ones being the Ito Calculus and the Stratonovic Calculus \cite{cwGardiner:2009a,vanKampen:2007a}.  This distinction vanishes for the Ornstein-Uhlenbeck process (OUP), which is the situation under consideration. Because our analysis is based on the use of  the Fokker-Planck diffusion equations, the stochastic calculus is not used in this note. } Our particular model with an  \emph{affine drift term}
\begin{align}
\alpha(X) = \alpha_0 - \alpha_1 X  = \mu - \kappa X , \quad \alpha_1 = \kappa  >0 , \label{eq:oupdrift}
\end{align}
and \emph{state-independent intensity},
\begin{align}
 \beta(X) = \beta_0 = \sigma^2 > 0 , \label{eq:oupintensity}
\end{align}
is known as the \emph{Ornstein-Uhlenbeck process} (OUP),
\begin{align}
dX(t) = \mu - \kappa\, X(t) + \sigma\,  dW(t) .  \label{eq:ouprocess}
\end{align}

\smallskip
\noindent If in addition a firing threshold with reset (after a possible refractory period) is enforced, we obtain the \emph{OUP Leaky integrate-and-fire model}, which we also refer to as the \emph{OUP/LIF neuron}.

\medskip
The LIF neuron fires when $V(t) = \theta$.  The first-passage time to threshold, $T_\theta$, beginning from a membrane voltage reset value $V(t_0) = v_0$ at initial time $t = 0$ is defined by,
\[ T_\theta =  T_\theta(v_0) =  \inf \{ t \, \big| \, V(t) = \theta, \, V(t_0) = v_0 < \theta \}  . \]
Note that under our model, $T$ is a random variable.  For the model under consideration in this note, i.e. the Ornstein-Uhlenbeck process (OUP), we desire to know if the neuron will fire in \emph{finite time} with probability one as well as the mean time to fire.  Note that if there is a finite probability that the neuron will never fire in finite time, then the mean time to fire must be infinite as $T = \infty$ will have a nonzero probability of occurring.  Thus, as noted by \cite{hcTuckwell:1988v2} one should always first ascertain that a neuron will fire (or not) in finite time with probability one.  As noted in the abstract (and discussed below) knowledge of the mean time-to-fire provides a measure of the firing rate of the neuron.  

\medskip
Although we have indicated, and will continue to indicate,  the dependence of the passage time $T$ on the threshold value $\theta$ and the 
initialization (i.e., the reset) voltage $v_0$, in fact we are interested in its dependence on the injected current $I_0$ which is our parameter of control to affect the firing behavior of the neuron, so we will write variously $T = T_\theta(v_0) =  T_\theta(v_0,I_0) = T(I_0)$, etc.

\medskip 
The goal of this note is to demonstrate that the OUP/LIF has a finite passage time with probability one, so that the neuron is almost surely guaranteed to fire in a finite time, and to derive its mean first passage  as a function of the injected current $I_0$, the reset voltage $V(t_0) = v_0$, and the threshold value $\theta$ which in turn allows us to determine the mean firing rate as a function of these quantities and the absolute refractory period $T_r$. At the price of some inelegance, the derivation is kept as simple has possible in order to be accessible to a wider range of students and nonspecialists.\footnote{In particular we avoid, or minimize, the discussion of moment generating functions; Laplace transforms and complex analysis; stochastic differential equations; the theory of the Sturm-Liouville problem; the theory of integral equations; and the theory of diffusion process boundary conditions, all of which have utility in a rigorous analysis of the first passage time problem \cite{lAlili:2005a,daDarling:1953a,nsGoel:1974a,piJohannesma:1968a,lmRicciardi:1977a,lmRicciardi:1990a,lSacerdote:2013a,ajSiegert:1951a}. Knowledge of the behavior of a first-order scalar stochastic dynamical system as described in the first 5 chapters of the textbook by Gardiner should provide an adequate background for following the development given in this note.} An attempt has been made to be thorough and detailed, and therefore many derivation steps are given which are often left out of the published literature and textbooks.


\section{SDE{\normalsize s}, FPE{\normalsize s} and the First-Passage Time}

\subsection{Stochastic Differential Equations and the Fokker-Planck Equation}
As motivated in the prior discussion, we are interested in the dynamical behavior of the homogeneous stochastic differential equation (SDE),
\begin{align}
\boxed{ \text{\bf SDE\, :} \qquad dX(t) = \alpha(X(t)) dt + \sqrt{ \beta(X(t))} \, dW(t)   \qquad t \ge 0 }  \label{eq:sde1}
\end{align}
where $W$ is a standard Wiener process.  With the assumption that the initial condition $X(0)$ is (possibly degenerate) gaussian and independent of $W$, this describes the behavior of a continuous sample path, homogeneous Markov process.\footnote{The process is homogeneous because the $\alpha(x)$ and $\beta(x)$ are independent of the time, $t$.}   Although we are primarily interested in the behavior of the OUP, we will keep the development fairly general and only specialize to the OUP towards the end of the note.

\medskip
Eq.~\eqref{eq:sde1} provides a local description of the behavior of a diffusion process.  There are many global solutions consistent with this local behavior, each one determined by the imposition of initial and boundary conditions.  
The Markov process \eqref{eq:sde1}  takes continuous values and therefore, once appropriate initial and boundary conditions have been imposed, has a behavior which is described by a transition probability density function, $p( x, t | x_0 , t_0)$, which uniquely satisfies the \emph{Chapman-Komolgorov equation} (CKE),\footnote{\label{ft:probdef} We define the conditional density function $p(x,t | x_0, t_0)$ as
\[ p(x,t \big| x_0, t_0) dx =  \text{Prob}\big\{ X(t) \in  (x, x + dx) \big| X(0) = x_0 \big\} , \]
with corresponding condition probability distribution,
\[ P(x, t \big| x_0, t_0) = \int_{- \infty}^x p(x',t \big| x_0, t_0) dx' = \text{Prob} \big\{ X(t) \le x \big| X(0) = x_0 \big\} . \]}
\begin{align}
\boxed{ \text{\bf CKE\, :} \qquad p( x, t | x_0 , t_0) = \int p(x,t | x', t') p(x',t' | x_0, t_0) dx' } \label{eq:CKE}
\end{align}
and the condition,
\[ p(x,t) = \int p(x,t | x', t') p(x',t') dx' , \]
for all admissible values of $x$ and $t$ \cite{cwGardiner:2009a}.

\medskip
Under mild assumptions on $\alpha(\cdot)$ and $\beta(\cdot)$ a distribution which satisfies the CKE also satisfies a differential version of the CKE, the (forward) \emph{ Fokker-Planck equation} (FPE)  \cite{lmRicciardi:1977a,vanKampen:2007a,cwGardiner:2009a},\footnote{Also known as the (forward) diffusion equation, the forward Kolmogorov equation, and the Smoluchowski equation.}$^,$\footnote{The equivalence between Eq.s~\eqref{eq:sde1} and \eqref{eq:FPE1} holds in general under the rules of the Ito stochastic calculus. In general the equivalence does not hold under the rules of the Stratonovich calculus.   Although most authors and researchers work primarily within the Ito formalism (e.g., \cite{cwGardiner:2009a,vanKampen:2007a}) some adhere to the Stratonovich formalism (e.g. \cite{nsGoel:1974a}), which can potentially cause confusion when comparing results across various source texts and papers. However, the Ornstein-Uhlenbeck process, which is the model under consideration in this note, fortuitously has the same behavior under both the Ito and Stratonovich interpretations \cite{cwGardiner:2009a}.}
\begin{align}
\frac{\partial}{\partial t} p( x, t | x_0 , t_0) = - \frac{\partial}{\partial x}\alpha(x)  p( x, t | x_0 , t_0) +  \frac{1}{2} \frac{\partial^2}{\partial x^2} \beta(x) p( x, t | x_0 , t_0) . \label{eq:FPE1}
\end{align}
There are many possible solutions, $f(x,t)$, to the Fokker-Planck equation,
\begin{align}
\boxed{ \text{\bf FPE\, :} \qquad \frac{\partial}{\partial t} f(x,t) = - \frac{\partial}{\partial x}\alpha(x)  f(x,t) +  \frac{1}{2} \frac{\partial^2}{\partial x^2} \beta(x) f(x,t)}\label{eq:FPE2}
\end{align}
and it is the imposition of initial conditions (at $t = t_0$) and boundary conditions that determines which particular solution is obtained.  In particular, if consistent sets of initial and boundary conditions are correctly chosen for the stochastic differential equation Eq.~\eqref{eq:sde1} and for  the FPE Eq.~\eqref{eq:FPE2}, the solution to Eq.~\eqref{eq:FPE2} will correctly describe the stochastic behavior of Eq.~\eqref{eq:sde1} and satisfy the corresponding CKE \eqref{eq:CKE}.  In particular the imposition of the initial condition (IC),
\begin{align}
 f(x,t_0) =  \delta(x - x_0), \label{eq:deltaic}  
\end{align}
the unrestricted boundary conditions,\footnote{It is possible to place alternative and/or additional conditions on the behavior of the solution $f(x,t)$ at $x = \pm \infty$, depending on the nature of the solution one is looking for \cite{nsGoel:1974a,cwGardiner:2009a,vanKampen:2007a}.}
\begin{align}
f(- \infty, t) = p(- \infty, t \big| x_0, t_0) = 0 \quad \text{and} \quad f(\infty, t) = p( \infty, t \big| x_0, t_0) = 0 , \label{eq:unrestrictedbc}.
\end{align} 
and the pdf normalization condition,\footnote{Note that the normalization condition is sufficient for the previous boundary conditions at $x = \pm \infty$.  One reason to decouple the two conditions is that under some alternative boundary conditions reasonable solutions can satisfy
\[ \int p(x, t \big| x_0, t_0) dx < 1, \]
where the loss of probability corresponds to the probability of ``aborption.'' For a discussion of this subtle point see \cite{vanKampen:2007a}.}
\[ \int p(x, t \big| x_0, t_0) dx = 1, \]
to the solution of the FPE \eqref{eq:FPE2} will result in a transition density $p(x,t \big| x_0, t_0)$ that correctly describes the stochastic behavior, $X(t)$, $t \ge 0$, of Eq.~\eqref{eq:sde1}, provided the latter has been subjected to the initial condition $X(t_0) = x_0$ with no other conditions imposed on $X(t)$.\footnote{This is called the \emph{unrestricted case} as no limitations are placed on the values taken by the state $X(t)$ for $t > t_0$. Note that the general solution involves $x_0$ and $t_0$ as ``free parameters.''}  On the other hand, if one applies the initial condition
$ f(x,t_0) = p(x,t_0)$, subject to the same boundary conditions, one obtains the solution $f(x,t) = p(x,t)$. Below, we will discuss other important solutions to the FPE, as well as to the backward FPE which is to be discussed next.

\medskip
Within the interior of an admissible  open state space region of the SDE~\eqref{eq:sde1}, the sample paths system are all continuous, provided that $\alpha(\cdot)$ and $\beta(\cdot)$ are reasonably well-behaved functions of their arguments.  This means that the solutions to the FPE, and its variants discussed below, will assumed to be continuous functions of their space and time arguments within the interior of the state space. Of course, once the state $X(t)$ hits the threshold value $\theta$, which is a \emph{boundary} of the state space interior, there is a radical, discontinuous reset of the state value, 
\[ X(\theta) \to X(t_0) = x_0 < \theta . \]

\subsection{The  Homogeneous Backward and Mixed Fokker-Planck Equations}

The density satisfying the CKE also satisfies the backward Fokker-Planck equation (BFPE),
\begin{align}
\boxed{ \text{\bf BFPE\, :} \quad - \frac{\partial}{\partial t_0} p( x, t | x_0 , t_0) = \alpha(x)  \frac{\partial}{\partial x_0}  p( x, t | x_0 , t_0) +  \frac{1}{2} \beta(x) \frac{\partial^2}{\partial x_0^2} p( x, t | x_0 , t_0) } \label{eq:BFPE1}
\end{align}
In this equation $x$ and $t$  are fixed (i.e., the final state and time are treated as parameters) and $x_0$ and $t_0$ are treated independent variables subject to the constraint $t_0 \le t$. If the SDE \eqref{eq:sde1} is subject to the initial condition $X(t_0) = x_0$ and is otherwise unrestricted, then its stochastic behavior is described by the solution to \eqref{eq:BFPE1} which satisfies the \emph{final condition} (FC),
\[ p(x,t | x_0, t) = \delta(x_0 - x) . \]

\medskip
This equation is often presented in a slightly different form in the homogeneous case which we are considering in this note.\footnote{It should be noted that a stationary process is necessarily homogenous.  (However, a homogenous process is not necessarily stationary, unless special initial conditions have been chosen or the process is considered in the asymptotic limit $t \to \infty$ given certain stability assumptions.)} Homogeneity means that for all $\tau$,
\[ p(x , t' + \tau | x_0, t'_0 + \tau) = p(x,t' | x_0, t'_0 ) . \]
In particular, for $\tau = - t'_0$, and elapsed time $t = t' - t'_0$, we have,
\[ p(x , t | x_0 , 0) = p(x , 0 | x_0 , -t) . \]
Taking making the assignments $t_0 = -t$ and $t=0$ in the left-hand-side of Eq.~\eqref{eq:BFPE1}, we have,
\[ - \frac{\partial}{\partial t_0} p( x, t | x_0 , t_0) \to - \frac{\partial}{\partial (-t) } p( x, 0 | x_0 , - t)  =  \frac{\partial}{\partial t} p( x, 0  | x_0 , - t )
=  \frac{\partial}{\partial t} p( x, t  | x_0 , 0 ) .\]
Thus if we define,
\[ g(y,t) = p(x,t | y, 0 ) , \]
the homogenous BFPE can be rewritten as the \emph{mixed} Fokker-Planck equation (MFPE), 
\begin{align}
 \boxed{ \text{\bf MFPE \, :} \qquad \frac{\partial}{\partial t} g(y,t) = \alpha(y)  \frac{\partial}{\partial y}  g( y, t ) +  \frac{1}{2} \beta(y) \frac{\partial^2}{\partial y^2} g( y, t ) } \label{eq:MFPE1}
\end{align}
Note that in Eq.~\eqref{eq:MFPE1} $t$ is the (``forward'') time which has elapsed since the initial time $t_0 = 0$ and $y$ is the value at the (``backward'') initial time, $X(0) = y$.\footnote{In terms of the notation used in Eq.~\eqref{eq:FPE1}, Eq.~\eqref{eq:MFPE1} is 
\[\frac{\partial}{\partial t} p(x,t | x_0 , 0) = \alpha(x_0)  \frac{\partial}{\partial x_0}  p(x,t | x_0 , 0) +  \frac{1}{2} \beta(x_0) \frac{\partial^2}{\partial x_0^2} p(x,t | x_0 , 0).\]} This variant of the BFPE can be found amply utilized in references  which primarily focus on the homogenous case (e.g., \cite{ajSiegert:1951a,daDarling:1953a,nsGoel:1974a}) as well as in other sources (see, e.g., Eq.~(5.5.5) of \cite{cwGardiner:2009a}).  

\medskip
Because Eq.~\eqref{eq:MFPE1}  has an independent parameter, $t$, that refers to the future (forward) time and an independent parameter that refers to the current (or backward) initial condition $y = X(0)$, we call Eq.~\eqref{eq:MFPE1}  the \emph{mixed} Fokker-Planck equation (MFPE)  in order to distinguish it from the (forward) FPE \eqref{eq:FPE2} and the BFPE~\eqref{eq:BFPE1}.  It is important to recognize the distinct differences between these three equations. 

\medskip
Under the appropriately consistent initial and boundary conditions the solutions to the FPE, BFPE, and MFPE are equal and describe the stochastic behavior of the SDE \eqref{eq:sde1}.  In particular, we are interested in the initial conditions  for the FPEs which correspond to $X(0) = y$,\footnote{Because of the assumption of homogeneity, henceforth, with no loss of generality, we take $t_0 = 0$, and henceforth we will always take $t \ge 0$.} 
 \begin{align}
 \boxed{ \text{\bf ICs \, :} \qquad f(x,0) = p(x,0|y, 0) = g(y, 0 ) = \delta(x - y) }  \label{eq:ics}
 \end{align}
As all three variants of the FPE yield the same solution for the transition density function, $p(x,t | y, 0)$,  we are free to use the one which is of greatest convenience given the nature of $\alpha(\cdot)$, $\beta(\cdot)$, and the boundary conditions. Note that the solutions to the FPE's obtained via the use of the ICs \eqref{eq:ics} have tacit dependencies; specifically $f(x,t)$ tacitly depends on $y$ while $g(y,t)$ tacitly depends on $x$.  In the sequel it will be useful to make these dependencies notationally overt via the definitions,
\[ f(x,t) = f_y (x,t) = f(x,t;y) \quad \text{and} \quad g(y,t) = g_x(y,t) = g(y,t;x) . \]

\subsection{The First Passage Time}

Consider the case where as long as $X(t) \in \mathsf{S}$, for some open, connected subset of the reals, $\mathsf{S}  \subset \mathbb{R}$, its dynamical behavior is described by the SDE \eqref{eq:sde1}.  If $\mathsf{S}= \mathbb{R}$, we say that the SDE is \emph{unrestricted,} whereas if $\mathsf{S}$ is a proper subset of the reals, we say that it is \emph{restricted.}  A standard choice is  an open internal, $\mathsf{S}_{a,\theta} = (a, \theta)$.  Our primary interest is with the case when $a \to - \infty$, 
\[ \mathsf{S} = \mathsf{S}_\theta = (-\infty, \theta). \]
Because the SDE \eqref{eq:sde1} models a diffusion process, it has continuous sample paths, and thus the only way that $X(t)$ can exit the admissible state space $\mathsf{S}$ is to hit the upper limit $\theta$ at some passage time\footnote{Also known as a hitting time, an exit time, an escape time, or a (gambler's) ruin time.} $t'$, $X(t') = \theta$. 

\medskip
In the leaky integrate-and-fire (LIF) neuron, as soon as (i.e., \emph{the first time that}) $X(t)$ ``escapes from'' the region $\mathsf{S}$, the neuron fires (an impulse is generated and sent to downstream neurons) and the dynamics represented by the SDE \eqref{eq:sde1} is ``turned off'' for a (deterministic) duration, $T_r$ chosen to model the absolute refractory period of the neuron, after which the clock is reset to zero, $t = t_0 = 0$, and the neuron is reset to its initialization value, $X(0) = x_0$. The process repeats itself indefinitely. 

\medskip
Once the neuron is reset the dynamical behavior \eqref{eq:sde1} is allowed to run in an unrestricted manner until the threshold $\theta$ is reached.  Thus one is interested in the \emph{first passage time}.  This is defined as,
\[T = T_\mathsf{S}(x_0) =  \inf \big\{ t \big| X(t) \not \in \mathsf{S}, \ X(0) = x_0  \in \mathsf{S} \big\} \]
For our situation of interest, we have 
\[T = T_\theta(x_0) = \inf \big\{ t \big| X(t) = \theta , \ X(0) = x_0 < \theta \big\} . \]

\medskip
The time between firings is given by the random variable $T_f = T_r+ T$ where $T$ is the random time which elapses between the ``unpinning'' of the membrane potential clamp and the next, subsequent firing of the neuron, as just defined. The \emph{mean time-to-fire} (MTF), 
\[ \widehat{T}_f = \text{E}(T_f) = T_r + \text{E}(T) = T_r + \widehat{T}, \]
 provides a measure $\rho$ of the  \emph{average firing rate} (AFR) of the neuron,
\[ \rho = \widehat{T}_f^{-1} = \frac{1}{T_r + \widehat{T}} . \]
The primary goal of this note is to derive the \emph{mean passage time} (MPT),
\[  \widehat{T} = \widehat{T}_\theta(x_0) = \text{E} \{ T_\theta(x_0) \} , \]
for the Ornstein-Uhlenbeck process (OUP) in a greater degree of detail than is easily found in the literature.

\subsection{The MFPE for the Restricted, Subthreshold Dynamics}

Because until the state $X(t)$ reaches the threshold value $\theta$ the local ``subthreshold'' dynamics are modeled by the SDE \eqref{eq:sde1},  its local, stochastic behavior until it escapes from the set $\mathsf{S}$ is completely captured by the FPEs.  Indeed, it's \emph{entire} stochastic behavior within the the interior of the region $\mathsf{S}$ is captured by the FPEs. What distinguishes the global solution to the FPEs for the restricted behavior ($X(t)$ confined to $\mathsf{S}$) from the global solution for the unrestricted behavior ($X(t)$ free to take any value in $\mathbb{R}$) are not the FPE's themselves (they after all describe the \emph{same} local motions \emph{within} the the interior of the set $\mathsf{S}$ for both the restricted and unrestricted cases) but rather the boundary conditions which are applied to the FPEs (because the difference between the two cases arises when the restricted motion bumps up against the boundary). 

\medskip
To handle the ``bumping up'' against the boundary $\theta$ which turns off the dynamics, we imagine that the SDE \eqref{eq:sde1} describes the behavior of a brownian motion particle that is absorbed (``vanishes'') as soon as it hits the position $\theta$, never to return again.\footnote{A nice discussion of this is given in \cite{vanKampen:2007a}.}  The theory of boundary conditions for Markov and diffusion processes is discussed at length in many places (see, e.g., \cite{nsGoel:1974a,cwGardiner:2009a}); suffice it to say that for our purposes it is enough to know that the appropriate boundary conditions to enforce in the FPE \eqref{eq:FPE2} and the MFPE \eqref{eq:MFPE1} if one is interested in solving for the transition density function are the respective absorbing boundary conditions,
\begin{align}
\boxed{ \text{\bf ABCs\, :} \qquad f(\theta, t) = p( \theta , t | x_0 , 0 ) =  0 \quad \text{and}  \quad g(\theta, t) =  p(x, t | \theta, 0) = 0 }  \label{eq:bcs}
\end{align}
The first condition says that the probability of being at $\theta$ is zero (because once you step there, you vanish) and the second condition says that once you've vanished, you never return to anywhere (because you've vanished for good).  Because a diffusion processes is continuous, these conditions imply that,
\begin{align}
 \text{Prob} \big\{ X(t) \ge \theta \big| X(0) = x_0  \big\} = 0 \quad \text{and} \quad  \text{Prob} \big\{ X(t)  \big| X(0) = x_0 \ge \theta  \big\} = 0 . \label{eq:bcs2}
\end{align}

\medskip
As described in a variety of sources, the BFPE is particularly useful for solving passage time problems \cite{nsGoel:1974a,lmRicciardi:1977a,hcTuckwell:1988v2,vanKampen:2007a,cwGardiner:2009a}, particularly in the guise of the MFPE variant given by Eq.~\eqref{eq:MFPE1} above \cite{ajSiegert:1951a,daDarling:1953a,nsGoel:1974a,cwGardiner:2009a}.\footnote{In particular, the development given in this note closely follows that beginning with Eq.~(5.5.5) on page 131 of \cite{cwGardiner:2009a}.}  
In particular, the solution of Eq.~\eqref{eq:MFPE1}, 
\begin{align}
\frac{\partial}{\partial t} g_x(y,t) = \alpha(y)  \frac{\partial}{\partial y}  g_x( y, t ) +  \frac{1}{2} \beta(y) \frac{\partial^2}{\partial y^2} g_x( y, t ) ,  \label{eq:restrictedfpe}
\end{align}
subject to the respective initial and boundary conditions, 
\begin{align} 
g_x(y,0) = \delta(y-x) \quad \text{and} \quad g_x(\theta, t) = 0 , \ \ \forall t \label{eq:restrictedbcs}
\end{align} 
is equal to the restricted dynamics probability density function, 
\[ g_x(y,t) = \widehat{p}(x,t |y,0), \]
where, 
\begin{align*} 
\widehat{p}(x, t |y, 0)dx  
= \text{Prob}\big\{ X(t) \in (x, x + dx), \ X(t') < \theta, \forall t' \in [0, t] \big| X(0) = y \big\} ,
\end{align*}
and
\begin{align*}
G_x(y,t) & = \widehat{P}(x|y,0) = \int_{-\infty}^x g_x(y,t)dx = \int_{-\infty}^x \widehat{p}(x,t | y, 0) dx \\
& = \text{Prob} \big\{ X(t) \le x, \ X(t') < \theta, \forall t' \in [0, t] \big| X(0) = y \big\} 
\end{align*}
is the probability distribution function for the restricted processes $X(t)$. Because Eq.~\eqref{eq:restrictedfpe} describes the behavior of a continuous sample path diffusion motion, the density function $g_x(y,t)$  and the distribution function $G_x(y,t)$ are  continuous functions of $x$, $y$ and $t$ within the interior of the state space. 

\subsection{Moments of the First Passage Time}

The development in this section closely follows that of Section 5.5.1 of \cite{cwGardiner:2009a}.  However, as noted in \cite{hcTuckwell:1988v2} there are some additional subtle points to be addressed in the derivation of the passage time moments.\footnote{Both \cite{nsGoel:1974a} and \cite{hcTuckwell:1988v2} stress the importance of ascertaining that the probability of a finite passage time be one in order to claim that a neuron will reliably fire when stimulated.  In addition, a probability-one finite mean passage time provides a technical justification for a seemingly minor, yet key, step made in the development given in \cite{cwGardiner:2009a}.}

\medskip
Note from the development in the previous section that,\footnote{Note that $G_\theta(y,0) = 1$ for all $y < \theta$ and that $G_\theta(y,t) = 0$ for $y \ge \theta$. In particular, $G_\theta(\theta,t) = 0$.}
\[ \text{Prob}\big\{ T = T_\theta(y) > t\big\} = G_\theta (y,t)  = \int_{-\infty}^\theta g_x(y,t) dx . \]
Thus the distribution function of $T$ is given by,\footnote{Note that $Q_\theta(y,t) = 1$ for all $y \ge \theta$ and that $Q_\theta(y,0) = 0$ as $X(0) = y < \theta$.}
\[ Q_\theta(y,t) =  \text{Prob}\big\{ T = T_\theta(y) \le t\big\} = 1 -  \text{Prob}\big\{ T = T_\theta(y) > t\big\} = 1- G_\theta(y,t) . \]
Having obtained the passage time distribution function $Q_\theta(y,t)$ we can compute the passage time probability density function as,\footnote{Note that $q_\theta(\theta,t) = 1$ and $q_\theta(y,0) = 0$ for $y < \theta$.}$^,$\footnote{For future reference note that $Q_\theta(y,t) = \int_0^t q_\theta(y,t) dt$ and in particular that $Q_\theta(y,\infty) = \int_0^\infty q_\theta(y,t) dt$, which should have a value of one if we are to have a finite passage time occur almost surely.}
\begin{align}
 q_\theta(y,t) = \frac{\partial}{\partial t} Q_\theta(y,t)  = - \frac{\partial}{\partial t} G_\theta(y,t) = - \frac{\partial}{\partial t} \int_{-\infty}^\theta g_x(y,t) dx . = - \int_{-\infty}^\theta \frac{\partial}{\partial t}  g_x(y,t) dx . \label{eq:qdef}
\end{align}
From the assumed continuity of $G_x(y,t)$ and $g_x(y,t)$, $Q_\theta(y,t)$ and $q_\theta(y,t)$ are similarly continuous. 

\medskip
Note that the condition, 
\begin{align}
Q_\theta(y,\infty) = \lim_{t \to \infty} Q_\theta(y, t) =1  \quad \forall y  \, ,  \label{eq:Qlim}
\end{align}
 if it holds, means precisely that the first passage time is finite, and therefore neuron will fire \emph{in finite time}, with probability one regardless of the initialization state.  Therefore we will spend some time to verify this fact for the OUP neuron. 
 
 \medskip
 
 What does it mean if $Q_\theta(y,\infty) <  1$?  It is a property of probability measures that they can be decomposed into a continuous part (like $Q_\theta(y, t)$), represented by continuous probability density functions, like
\[ q_\theta(y,t) = \frac{\partial}{\partial t} Q_\theta(y, t) , \]
which have no jumps, and a discrete, ``atomic'' part that places nonzero probability mass on discrete points \cite{aPapoulis2002a}.\footnote{There is also a third, mathematically abstract,``singular'' part which is not encountered when dealing with phenomena in the physical world.} If $Q(_\theta(y,\infty) = 1$, then $Q(_\theta(y,t)$ is a distribution which describes a completely continuous random variable, $T$, that with probability one takes a finite value $T= t$, $0 < T < \infty$. On the other hand, if $Q_\theta(y, \infty) < 1$, then \emph{with probability}, 
\[ G_\theta(y,\infty) = 1 - Q_\theta(y, \infty), \]
 the threshold $\theta$ will not be hit in any \emph{finite time}, which means that there must be a discrete, atomic amount (equal to $1 - Q_\theta(y, \infty)$) of probability mass (conceptually) located at $t = \infty$. In this latter case there is a nonzero probability that the neuron will \emph{never} fire.\footnote{For an example of a model neuron that has a nonzero probability of never firing, see the example on page 142 of \cite{hcTuckwell:1988v2}.}

\medskip
We do assume that, for each $y$, $\lim_{t \to \infty} Q_\theta(y, t) = Q_\theta(y,\infty)$ converges to a constant value that is independent of $t$.\footnote{We prove later that $Q_\theta(y,\infty) \le 1$.}  And if
\[ Q_\theta(y,\infty) = \int_0^\infty q_\theta(y,t) dt \] 
converges, it must be the case that,\footnote{Thus we have that $q_\theta(y,0) = q_\theta(y,\infty) = 0$ for $y < \theta$. We can have $q_\theta(y,\infty) = 0$  and yet have a nonzero probability mass at $t = \infty$ because $q_\theta(y,t)$ represents the purely continuous part of the $T$-distribution. It is unfortunately the case that egregiously using $q_\theta(y,\infty)$ in place of the more accurate $\lim_{t \to \infty} q_\theta(y,t)$ can be misleading: $q_\theta(y,\infty)$ does \emph{not} represent probability \emph{mass at infinity}, but rather the \emph{probability 
density on the way to infinity}.}
\begin{align}
q_\theta(y, \infty) = 0 . \label{eq:aatinf}
\end{align}

 \medskip
 Note that Eq.~\eqref{eq:Qlim} is true if and only if
\begin{align}
 G_\theta(y,\infty) = \lim_{t \to \infty} G_\theta(y, t) = 0, \quad \forall y \, . \label{eq:gtozero}
\end{align}
Although condition \eqref{eq:gtozero}  is often assumed to be true,\footnote{For example, see Equation (5.5.16) of \cite{cwGardiner:2009a}.} as noted in \cite{hcTuckwell:1988v2} this is a key property\footnote{Equivalent to the probability one finite passage time condition \eqref{eq:Qlim}.} that ideally should be verified. Of course, \eqref{eq:gtozero} can always just be assumed to be true.\footnote{And then hopefully verified after the fact, thereby demonstrating that a self-consistent derivation has occurred.}  Indeed, we will do just that by (temporarily) assuming a stronger condition, namely that the convergence of $G_\theta(y,t)$ obeys,
\begin{align}
\lim_{t \to \infty} t^n G_\theta(y, t) = 0 \, ,\quad \forall y \,  , \label{eq:tgtozero}
\end{align} 
for specific values of $n \ge 0$.
When such assumptions are made, we will be very careful to note them.\footnote{Eventually we will replace the condition \eqref{eq:tgtozero} with the finite moments condition \eqref{eq:finitemoments}.}

\medskip
To demonstrate an application of the condition \eqref{eq:tgtozero}, we first define the moments of the first passage time random variable $T = T_\theta(y)$ by 
\begin{align} m_n = m_n(\theta,y) = \text{E} \big\{ T^n \} = \int_{0}^\infty t^n  q_\theta(y,t) dt =  - \int_{0}^\infty t^n  \frac{\partial}{\partial t} G_\theta(y,t) dt.  \label{eq:momentsdef}
\end{align}
Note, in particular, that the ``zeroth moment'', $m_0$, has already been encountered,
\[  m_0(\theta,y) = Q_\theta(y,\infty)  = 1 - G_\theta(y,\infty) , \]
and gives the total probability that the first passage time from the initial state $y$ is finite.  As we have noted, 
\[ m_0(\theta,y) = 1 \  \ \text{for all} \  \ y \quad \iff \quad  G_\theta(y,\infty) =0  \ \ \text{for all} \ \ y. \]

\medskip
To compute higher order moments, $m_n(\theta,y)$ from Eq.~\eqref{eq:momentsdef}, it is useful to note that,
\[ \frac{\partial}{\partial t} t^n G_\theta(y,t) dt =  t^n \frac{\partial}{\partial t} G_\theta(y,t) dt + n \,  t^{n-1} G_\theta(y,t) dt \]
so that integrating the last integral on the far right hand side of \eqref{eq:momentsdef} by parts yields
\begin{align}
m_n(\theta,y)  =  n \int_{0}^\infty t^{n-1} G_\theta(y,t) dt - \lim_{t\to \infty} t^n G_\theta(y,t)  +  \lim_{t\to 0} t^n G_\theta(y,t).  
\end{align}
For $n = 0$, this yields,
\[ m_0(\theta,y) = G_\theta(y,0) - G_\theta(y,\infty) = 1 - G_\theta(y,\infty), \]
as expected. For $n \ge 1$, we have,
\begin{align}
m_n(\theta,y)  =  n \int_{0}^\infty t^{n-1} G_\theta(y,t) dt - \lim_{t\to \infty} t^n G_\theta(y,t), \label{eq:mng1}
\end{align}
which, in turn, gives
\[ m_n(\theta,\theta)  = 0 \quad \text{for all $t \ge 1$}. \]
Finally, note that by invoking condition \eqref{eq:tgtozero},  Eq.~\eqref{eq:mng1} becomes the very useful relationship,
\begin{align}
\text{Condition \eqref{eq:tgtozero}} \implies m_n(\theta,y)  =  n \int_{0}^\infty t^{n-1} G_\theta(y,t) dt \quad \text{for $n \ge 1$}, \label{eq:mident}
\end{align}

\medskip
Inspired by Eq.~\eqref{eq:qdef}, integrate every term in Eq.~\eqref{eq:restrictedfpe} by $- \int_{-\infty}^\theta (\, \cdot \, ) dx$ to obtain the equation,
\begin{align} 
q_\theta (y,t) = - \alpha(y)  \frac{\partial}{\partial y}  G_\theta( y, t ) - \frac{1}{2} \beta(y) \frac{\partial^2}{\partial y^2} G_\theta ( y, t ) .  \label{eq:qeq}
\end{align}
Now differentiate every term by $t$,
\begin{align} 
 \frac{\partial}{\partial t} q_\theta (y,t) = - \alpha(y)  \frac{\partial}{\partial y}  \frac{\partial}{\partial t}  G_\theta( y, t ) - \frac{1}{2} \beta(y) \frac{\partial^2}{\partial y^2} \frac{\partial}{\partial t}  G_\theta ( y, t ) . \label{eq:firstmzero}
\end{align}
Finally, integrate every term by $\int_0^\infty (\, \cdot \, ) dt$,
\[ \underbrace{\int_0^\infty \frac{\partial}{\partial t} q_\theta (y,t) dt}_{q_\theta (y,\infty)  - q_\theta (y,0) = 0}
  = \alpha(y)  \frac{\partial}{\partial y}  \underbrace{\int_0^\infty \underbrace{\left( - \frac{\partial}{\partial t} G_\theta( y, t ) \right)}_{q_\theta(y,t)} dt}_{m_0(\theta,y)} +  \frac{1}{2} \beta(y) \frac{\partial^2}{\partial y^2} \underbrace{\int_0^\infty \underbrace{ \left( - \frac{\partial}{\partial t} G_\theta ( y, t ) \right)}_{q_\theta(y,t)} dt}_{m_0(\theta,y)} ,\] 
to obtain,
\begin{align} 
0 =  \alpha(y)  \frac{\partial}{\partial y}  \, m_0(\theta,y) +  \frac{1}{2} \beta(y) \frac{\partial^2}{\partial y^2} \, m_0(\theta,y) . \label{eq:mzero}
\end{align} 
Eq.~\eqref{eq:mzero} is to be solved for the finite passage time probability subject to the boundary conditions,
\begin{align}
\lim_{y \to - \infty} m_0 (\theta, y) =  m_0(\theta, - \infty) = 0 \quad \text{and} \quad m_0(\theta, \theta) = 1 . \label{eq:mzerobcs}
 \end{align}
Setting,
\[ m_0'(\theta,y) = \frac{\partial}{\partial y}  \, m_0(\theta,y)  , \]
we have,
\[\frac{\partial}{\partial y}  \, m'_0(\theta,y) = - 2 \frac{\alpha(y)}{\beta(y)}  m'_0(\theta,y) , \]
or
\[ m'_0(\theta,y) = \lim_{\ell \to - \infty} C_0 \exp \left\{ - 2 \int_{\ell}^y \frac{\alpha(y')}{\beta(y')}dy' \right\} . \]
Integrating one more time,\footnote{Differential equations that can be solved in this manner, i.e., via successive integrations, are said to be ``solvable by quadrature.''}
\[ m_0(\theta,y) =  C_1 + \lim_{\ell \to - \infty} C_0 \int_{\ell}^y \exp \left\{ - 2 \int_{\ell}^{y''} \frac{\alpha(y')}{\beta(y')}dy' \right\} dy'' . \]
Invoking the boundary conditions \eqref{eq:mzerobcs} yields the answer,
\begin{align}
m_0(\theta,y) = \lim_{\ell \to - \infty} \frac{\int\limits_{\ell}^y \exp \left\{ - 2 \int\limits_{\ell}^{y''} \frac{\alpha(y')}{\beta(y')}dy' \right\} dy''}{\int\limits_{\ell}^\theta \exp \left\{ - 2 \int\limits_{\ell}^{y''} \frac{\alpha(y')}{\beta(y')}dy' \right\} dy''}\le 1  \qquad \text{for} \quad y \le \theta . \label{eq:probonecond}
\end{align}
Note we have verified the claim made earlier that $m_0(\theta,y) = Q_\theta(y,\infty) \le 1$. If is apparent that  if $\alpha(y)$ and $\beta(y)$ are such that the limit of the right hand side of Eq.~\eqref{eq:probonecond} is one, then,\footnote{Note that the right hand side verifies \eqref{eq:tgtozero} for $n=0$.}
\begin{align}
m_0(\theta,y) = Q_\theta(y,\infty) = 1 \iff G_\theta(y,\infty) = 0 \label{eq:mzerotoone}
\end{align}
for all $y$, and then with probability one the first passage time is finite.  We will see below that this is true for the OUP.\footnote{That this holds for the OUP is also discussed on page 171 of \cite{hcTuckwell:1988v2}.}  

\medskip
Now assume that condition \eqref{eq:tgtozero} holds for $n=1$,
\[ \lim_{t \to \infty} t \,  G_\theta(y, t) = 0, \quad \forall y  . \]
Then \eqref{eq:mident} holds for $n=1$, yielding the following expression for the mean first passage time,
\begin{align}
 \text{E} \big\{ T_\theta(y)  \big\} = m_1(\theta,y)  =  \int_{0}^\infty G_\theta(y,t) dt . \label{eq:moneid}
\end{align}
Next integrate all terms in Eq.~\eqref{eq:qeq} by $\int_0^\infty (\, \cdot \, ) dt$ 
to obtain,
\begin{align*} 
- \int_0^\infty q_\theta (y,t) dt =  \alpha(y)  \frac{\partial}{\partial y}  \int_0^\infty G_\theta( y, t ) dt + \frac{1}{2} \beta(y) \frac{\partial^2}{\partial y^2} \int_0^\infty G_\theta ( y, t ) dt. 
\end{align*}
Finally by invoking  \eqref{eq:mzerotoone} and \eqref{eq:moneid}  we have the differential equation,  
\begin{align} 
- 1 = \alpha(y)  \frac{\partial}{\partial y}  m_1(\theta,y) + \frac{1}{2} \beta(y) \frac{\partial^2}{\partial y^2} m_1(\theta,y), \label{eq:onemzero}
\end{align}
which can be solved for the first moment $m_1(\theta,y)$ subject to the boundary condition,
\[ m_1(\theta,\theta ) = 0 . \]

\medskip
Equations  \eqref{eq:mzero} and \eqref{eq:onemzero} are special cases of the,\footnote{This recursion is given as Eq.~(6.5) in \cite{daDarling:1953a}. For now it assumes that condition \eqref{eq:tgtozero} holds, though this will be slightly weakened below and then replaced by the sufficient condition that the moments are assumed to exist.} 

\bigskip
\noindent{\bf \qquad  \qquad  \qquad \quad \quad \ \ \ \  \ \footnotesize DARLING \& SIEGERT MOMENTS RECURSION \cite{daDarling:1953a}} 
\begin{align} 
 \boxed{ -  n  \, m_{n-1} (\theta,y)  = \alpha(y)  \frac{\partial}{\partial y}  m_n (\theta,y) + \frac{1}{2} \beta(y) \frac{\partial^2}{\partial y^2} m_n (\theta,y) }\label{eq:nmzero} 
\end{align}
which is to be solved subject to the conditions,
\[ m_{-1}(\theta,y) = 0, \quad m_0(\theta,\theta) = 1 \quad \text{and} \quad m_n (\theta,\theta) = 0 \quad \text{for} \quad n \ge 1 . \]
The recursion \eqref{eq:nmzero} is straightforward to derive for  $n \ge 1$,\footnote{The case $n=0$ has been handled separately above.} assuming, for now, that \eqref{eq:tgtozero} holds for the desired values of $n$.  This is done by integrating all terms in Eq.~\eqref{eq:qeq} by $ - n \int_0^\infty  t^{n-1} (\, \cdot \, ) dt$, to obtain,
\[  - n \int_0^\infty  t^{n -1} q_\theta (y,t)   dt = \alpha(y)  \frac{\partial}{\partial y} \left( n \int_0^\infty  t^{n-1} G_\theta( y, t )  dt \right) + \frac{1}{2} \beta(y) \frac{\partial^2}{\partial y^2} \left(n \int_0^\infty  t^{n -1} G_\theta ( y, t )  dt  \right),\]
which from Eq.~\eqref{eq:momentsdef} and \eqref{eq:mident} immediately yields the moments recursion \eqref{eq:nmzero}.

\medskip
There is an alternative derivation of \eqref{eq:nmzero} that will allow us to replace the assumption that \eqref{eq:tgtozero} holds  with the assumption of the existence of the passage time moments.\footnote{The assumption of finite moments is the standard one encountered in the literature \cite{daDarling:1953a}.  As mentioned previously assumption \eqref{eq:tgtozero} is used in \cite{cwGardiner:2009a}.}  Because,
\[ q_\theta(y,t) =  - \frac{\partial}{\partial t} G_\theta(y, t) ,  \] 
we have that Eq.~\eqref{eq:firstmzero}  yields the \emph{first passage time diffusion equation} (FPTDE)  for the first passage time probability density function, 

\medskip
{\bf \qquad \quad \quad  \quad \quad \quad \ \ \, \ \,  \, FPTDE} 
\begin{align} 
\boxed{ \begin{aligned} \frac{\partial}{\partial t} q_\theta (y,t) & = \alpha(y)  \frac{\partial}{\partial y} \,   q_\theta( y, t ) + \frac{1}{2} \beta(y) \frac{\partial^2}{\partial y^2}  \, q_\theta ( y, t )  \label{eq:qdiffusion} \\
 \text{subject to} & \ \ q_\theta(y,0) = 0 \ \ \text{and} \ \ q_\theta(\theta,t) = 1 \, 
\end{aligned}}
\end{align}
If we now integrate every term in this diffusion equation against $t^n$,\footnote{I.e., by $\int_{0}^{\infty} t^n (\, \cdot \, ) dt$.} we get,
\[  \int_0^\infty  t^n \frac{\partial}{\partial t} q_\theta (y,t) dt  = \alpha(y)  \frac{\partial}{\partial y}  \, m_n(y) + \frac{1}{2} \beta(y) \frac{\partial^2}{\partial y^2} \,  m_n ( y ) .\]
Then, integration by parts yields,
\[ \int_0^\infty  t^n \frac{\partial}{\partial t} q_\theta (y,t) dt = -  n \underbrace{\int_0^\infty t^{n-1}  q_\theta (y,t) dt}_{m_{n-1}(\theta,y)} + \lim_{t \to \infty} t^n q_\theta (y,t)  , \]
so that,
\[  - n \, m_{n-1}(\theta,y) +   \lim_{t \to \infty} t^n q_\theta (y,t)  = \alpha(y)  \frac{\partial}{\partial y}  \, m_n(y) + \frac{1}{2} \beta(y) \frac{\partial^2}{\partial y^2} \, m_n ( y )  .\]
It is immediately apparent that the Darling \& Siebert Moments Recursion \eqref{eq:nmzero} holds if we assume that,
\begin{align}
\lim_{t \to \infty} t^n q_\theta (y,t) =  - \lim_{t \to \infty} t^n \frac{\partial}{\partial t} G_\theta(y, t)  = 0 ,  \label{eq:tPgtozero} 
\end{align}
a condition which should be compared to condition \eqref{eq:tgtozero}.  Condition \eqref{eq:tPgtozero} is equivalent to the condition,
\[ \lim_{t \to \infty} t^{n-1} G_\theta(y, t)  = 0 , \]
which is actually weaker than condition \eqref{eq:tgtozero}.\footnote{It is weaker because for the moments recursion \eqref{eq:nmzero} to hold for the $n$-th moment on the right hand side, it is now sufficient for \eqref{eq:tgtozero} to hold for only $n-1$.}  This is a consequence of L'H\^{o}pital's rule,
\[ 0 = \lim_{t \to \infty} t^{n-1} G_\theta(y, t) = \lim_{t \to \infty} \frac{G_\theta(y, t)}{t^{1-n}} 
= \lim_{t \to \infty} \frac{\partial G_\theta(y, t)/\partial t}{(1-n) t^{-n}}  =\frac{1}{1-n}  \lim_{t \to \infty} t^n \frac{\partial}{\partial t} G_\theta(y, t) . \]
A sufficient condition for Eq.~\eqref{eq:tPgtozero} to hold, is that the $n$-th first passage time moment  exist,
\[ m_n(\theta,y) = \int_0^\infty t^n q_\theta(y,t) dt < \infty . \]
We have thus demonstrated the standard sufficient condition for recursion \eqref{eq:nmzero}:

\begin{framed}
\noindent {\bf \ Sufficient Condition for the Darling \& Siegert Moments Recursion  to hold  \cite{daDarling:1953a}}

\medskip
\begin{quote}
Assume  that $m_0(\theta,y) = 1$. If all first passage time moments exist for $n \le n_0$, 
\begin{align}
m_n(\theta,y) < \infty,  \label{eq:finitemoments}
\end{align}
then the Darling \& Siegert Moments Recursion \eqref{eq:nmzero} is  valid for $n \le n_0$.
\end{quote}
\end{framed}

To actually verify condition \eqref{eq:finitemoments} in an  \emph{a priori} manner is usually nontrivial.  It involves analytical and/or numerical knowledge of $q_\theta(y,t)$ which is obtained from solving the diffusion equation \eqref{eq:qdiffusion}. This is generally a difficult task \cite{lAlili:2005a,lSacerdote:2013a}. Because of the difficulty in verifying \eqref{eq:finitemoments}, we will henceforth assume that all  first passage time $n$--th degree moments, $n \ge 1$, of interest exist.

\medskip
A necessary condition that moments exist for $n \ge 1$ is that $m_0(\theta,y) = 1$.  This is because if $m_0(\theta,y) < 1$, there must be nonzero probability mass at infinity which would case all moments $n \ge 1$ to be infinite.\footnote{This is discussed in \cite{hcTuckwell:1988v2}.} So ideally, at the very least, one should verify that $m_0 (\theta,y) = 1$.  We do this for the OUP model in the next subsection.

\medskip
We can solve Eq.~\eqref{eq:nmzero} to obtain a recursive formula expressing the moment $m_n(\theta,y)$ in terms of the lower order moment $m_{n-1}(\theta,y)$.  To do so let $m_n(y) = m_n(\theta,y)$ and define,
\[ m'_n(y) = \frac{\partial}{\partial y} m_n(y), \quad a(y) = 2 \frac{\alpha(y)}{\beta(y)} \quad \text{and} \quad b(y) = - 2 n\,  \frac{m_{n-1}(y)}{\beta(y)} . \]
Then (for fixed $\theta$) Eq.~\eqref{eq:nmzero}  becomes,
\begin{align}
 \frac{d}{d y} m'_n(y) + a(y) m'_n(y)  = b(y) . \label{eq:firstorder}
\end{align} 
This is a linear, inhomogenous first-order differential equation that can be solved by the standard method of variation of parameters. The general solution is the sum of a particular solution, $\psi_p(y)$, and a homogeneous solution, $C_0 \, \psi(y)$, where $C_0$ is a constant to be determined.  For the homogeneous case, we have,
\[ \frac{d}{d y} \psi(y) + a(y) \psi(y)  = 0 ,\]
which is solved by,\footnote{Similarly to the derivation to $m_0(y)$ done above, one more carefully would take the lower limit of integration to be $\ell$ and then take $\ell \to - \infty$ to deal with terms involving nonconvergent integrals.   We avoid this fussiness for now, but later we will have to bring the parameter $\ell$ into the foreground in order to obtain sensible answers.} 
\[ \psi(y) = \exp\left\{- \int_{- \infty}^y a(y') dy' \right\} =  \exp\left\{  - 2 \int_{-  \infty}^y \frac{\alpha(y')}{\beta(y')}  dy' \right\} . \]
A particular solution is then given by,
\[ \psi_p(y) = \psi(y) \int_{- \infty}^y \frac{b(y')}{\psi(y')} dy' , \]
as can be readily checked.  Thus, the general solution for $m'_n(y)$ is,
\[ m'_n(y) = C_0\,  \psi(y) + \psi_p(y) . \]
Another integration yields,
\begin{align*}
m_n(y) 
& = \int_{-\infty}^y\left(  C_0\,  \psi(y'') + \psi_p(y'') \right) d y'' + C_1 \\ 
& = \int_{-\infty}^y\left(  C_0\,  \exp\left\{ \int_{- \infty}^{y''} a(y') dy' \right\}+ \psi(y'') \int_{- \infty}^{y''}\frac{b(y')}{\psi(y')} dy' \right) d y'' + C_1 
\end{align*}
The boundary condition $m_n(\theta) = 0$ leads to the choices $C_0 = 0$ and,
\[ C_1 =  - \int_{-\infty}^\theta \left(  \psi(y'') \int_{- \infty}^{y''}\frac{b(y')}{\psi(y')} dy' \right) d y'' . \]
Then (note that $y \le \theta$),
\[ m_n(y) = - \int_y^{\theta}  \psi(y'')\left( \int_{- \infty}^{y''}\frac{b(y')}{\psi(y')} dy' \right) d y'' . \]
Thus we have shown that,
\begin{align}
\begin{aligned}
m_n(y) & = 2n \int_y^{\theta}  \psi(y'')\left( \int_{- \infty}^{y''}\frac{\, m_{n-1}(y')}{\beta(y') \psi(y')} \, dy' \right) d y'' \label{eq:siegert} \\
\text{where} & \quad \psi(y) =  \exp\left\{  - 2 \int_{-  \infty}^y \frac{\alpha(y')}{\beta(y')}  dy' \right\} . 
\end{aligned}
\end{align} 

\medskip
In the literature rather than working with $\psi(y)$, it is more common to see results written in terms of
\[ \psi^{-1}(y) = \exp\left\{  2 \int_{-  \infty}^y \frac{\alpha(y')}{\beta(y')}  dy'\right\} . \]
For this reason we define,\footnote{The quantity $\mathsf{W}(y)$ also has an important interpretation. it is often the case that $C$ can be chosen so that $\mathsf{W}(y)$ is normalized to be a probability density function.  In this case it is not hard to show that $W(x)$ is then the steady-state ($t \to \infty$) probability density function which is the stationary solution to the non-thresholded (forward) Fokker-Planck Equation \eqref{eq:FPE2},\[  - \frac{\partial}{\partial x}\alpha(x)  \mathsf{W}(x) +  \frac{1}{2} \frac{\partial^2}{\partial x^2} \beta(x) \mathsf{W}(x) = 0  \]
\[ \mathsf{W}(x) = \lim_{t \to \infty}  p( x, t | x_0 , t_0) . \]  The derivation of $W(x)$ as a solution to the steady-state Fokker-Planck equation is given in the Appendix.}
\begin{align}
\mathsf{W}(y)  = \frac{C}{\beta(y)} \psi^{-1} (y) =  \frac{C}{\beta(y)} \exp\left\{  2 \int_{-  \infty}^y \frac{\alpha(y')}{\beta(y')}  dy'\right\} . \label{eq:W1}
\end{align}

Notice that the inclusion of the constant $C$ in the definition of $\mathsf{W}(y)$ does not affect the formula in Eq.~\eqref{eq:siegert}, and thus we can rewrite \eqref{eq:siegert} as the,

\bigskip
{\bf \qquad \quad \ \ \  \  \quad \quad \, \ \ \ \ SIEGERT FORMULA \cite{ajSiegert:1951a}} 
\begin{align}
\boxed{\begin{aligned}
m_n(y) & = 2n \int_y^{\theta}  \frac{d y''}{\beta(y'')\mathsf{W}(y'')} \left( \int_{- \infty}^{y''} \mathsf{W}(y') m_{n-1}(y') \, dy' \right) \label{eq:siegert2} \\
\text{with} & \quad \mathsf{W}(y) =  \frac{C}{\beta(y)}  \exp\left\{  2 \int_{-  \infty}^y \frac{\alpha(y')}{\beta(y')}  dy' \right\} . 
\end{aligned}}
\end{align}
This important relationship  between consecutive moments of the passage time is  referred to in the literature as the {Siegert formula} \cite{lmRicciardi:1990a,lSacerdote:2013a} and is given as Equation (3.14) in Siegert's original 1951 paper \cite{ajSiegert:1951a}.\footnote{This formula also appears to have been independently discovered by Johannesma (see Eq.~(15) in \cite{piJohannesma:1968a}). Another early reference to the Siegert formula in the context of neuronal modeling is given in \cite{rmCapocelli:1971a} (see Eq.~(5.1) in this latter reference).}

\subsection{Mean First Passage Time for the OUP}

We now use the theory developed in the previous section to show that the first passage time is finite with probability one and to compute the mean first passage time for the Ornstein-Uhlenbeck Process (OUP), which is given by the SDE \eqref{eq:sde1} with,
\[ \alpha(x) = \mu - \kappa x , \ \kappa > 0, \ \  \text{and} \ \  \beta(x) = \sigma^2 > 0 , \]
so that,
\[ dX = \left( \mu - \kappa x \right) dt + \sigma\, dW . \]

As mentioned earlier dealing with the limit $y \to - \infty$ is a touchy issue. For this reason we proceed by setting the lower limits in the integrals of the Siegert formula which are at $- \infty$ to the finite value $\ell$ and after having formed the final quantities of interest we will take the limit $\ell \to - \infty$.\footnote{The rigorous way to handle this case is to define the two boundary diffusion problem, $\ell < x < \theta$, with appropriate boundary conditions at $\ell$ and $\theta$.  After the first passage time problem has been rigorously solve for this case, one can take $\ell \to - \infty$ \cite{nsGoel:1974a,lmRicciardi:1977a,vanKampen:2007a,cwGardiner:2009a}.  Rather than have to devote time to discuss the somewhat subtle issue of diffusion boundary conditions, in this note we have opted to work with improper integrals and evaluate expressions containing such integrals as  appropriately defined limiting cases.}  Thus we will proceed via the steps,
\[ \text{function of} \  \int\limits_{-\infty} ( \, \cdot \,) \, dy \to  \text{function of} \ \int\limits_{\ell } ( \, \cdot \,) \, dy \to \lim_{\ell \to - \infty} \ \text{function of} \int\limits_{\ell } ( \, \cdot \,) \, dy . \]

We begin by computing,
\begin{align*}
 2 \int\limits_{\ell}^y \frac{\alpha(y')}{\beta(y')}  dy'  & = \frac{2}{\sigma^2} \int\limits_\ell^y (\mu - \kappa y') dy'  \\
 & = \frac{2}{\sigma^2} \int\limits_0 ^y (\mu - \kappa y') dy' + \frac{2}{\sigma^2} \int\limits_\ell^0 (\mu - \kappa y') dy'\\
 & =  - \frac{\kappa}{\sigma^2} \left(y^2 - 2 \, \frac{\mu}{\kappa} \, y  \right) + \frac{\kappa}{\sigma^2} \left(\ell^2 - 2 \, \frac{\mu}{\kappa} \, \ell  \right)\\
  & =  - \frac{\kappa}{\sigma^2} \left(y^2 - 2 \, \frac{\mu}{\kappa} \, y + \frac{\mu^2}{\kappa^2} \right) + \frac{\mu^2}{\kappa \sigma^2} - \frac{\mu^2}{\kappa \sigma^2}  + \frac{\kappa}{\sigma^2} \left(\ell^2 - 2 \, \frac{\mu}{\kappa} \, \ell + \frac{\mu^2}{\kappa^2} \right) \\
  & =  - \frac{\kappa}{\sigma^2} \left(y - \frac{\mu}{\kappa} \right)^2 + \frac{\kappa}{\sigma^2} \left(\ell - \frac{\mu}{\kappa} \right)^2.
 \end{align*}
This shows that,
\begin{align}
\exp\left\{ 2 \int\limits_{\ell}^y \frac{\alpha(y')}{\beta(y')}  dy' \right\}  
= \exp\left\{- \frac{\kappa}{\sigma^2} \left( y - \frac{\mu}{\kappa} \right)^2 \right\}  
   \exp\left\{  \frac{\kappa}{\sigma^2} \left(\ell - \frac{\mu}{\kappa} \right)^2\right\} , \label{eq:exp1}
  \end{align}
   and 
 \begin{align}
 \exp\left\{ - 2 \int\limits_{\ell}^y \frac{\alpha(y')}{\beta(y')}  dy' \right\}  
= \exp\left\{ \frac{\kappa}{\sigma^2} \left( y - \frac{\mu}{\kappa} \right)^2\right\}  
   \exp\left\{ -  \frac{\kappa}{\sigma^2} \left(\ell - \frac{\mu}{\kappa} \right)^2\right\} \label{eq:exp2} .
   \end{align}
   
The latter expression allows Eq.~\eqref{eq:probonecond} to be written as,
 \begin{align}
m_0(\theta,y) = \lim_{\ell \to - \infty} \frac{\int\limits_{\ell}^y \exp \left\{  \frac{\kappa}{\sigma^2} \left( y'' - \frac{\mu}{\kappa} \right)^2\right\} dy''}{\int\limits_{\ell}^\theta \exp \left\{  \frac{\kappa}{\sigma^2} \left( y'' - \frac{\mu}{\kappa} \right)^2 \right\} dy''}  = 1 , \quad \text{for all $y \le \theta$},
\end{align}
showing that the probability of the OUP neuron firing in a finite time is one.\footnote{See also Eq.~(9.237) of \cite{hcTuckwell:1988v2}. Note that we have satisfied the necessary condition that all higher order moments exist.}

\medskip
Having shown that $m_0(y) \equiv 1$, from the Siegert formula \eqref{eq:siegert2} and Eq.~\eqref{eq:exp1} we can compute $m_1(y)$,\footnote{Assuming that it exists, $m_1(y) < \infty$.} 
\begin{align*}
m_1(y)  & = 2 \int_y^{\theta}  \frac{d y''}{\beta(y'')\mathsf{W}(y'')} \left( \int_{- \infty}^{y''} \mathsf{W}(y') \, dy' \right) \\
& = \lim_{\ell \to - \infty} 2 \int_y^{\theta}  \frac{d y''}{\sigma^2 \exp\left\{ - 2 \int\limits_{\ell}^{y''} \frac{\alpha(y')}{\beta(y')}  dy' \right\} } \left( \int_{- \infty}^{y''} \exp\left\{ - 2 \int\limits_{\ell}^{y'} \frac{\alpha(y)}{\beta(y)}  dy \right\}  \, dy' \right) \\
& = 2 \int_y^{\theta}  \frac{d y''}{\sigma^2 \exp\left\{ - \frac{\kappa}{\sigma^2} \left( y'' - \frac{\mu}{\kappa} \right)^2\right\}  } \left( \int_{- \infty}^{y''} \exp\left\{-  \frac{\kappa}{\sigma^2} \left( y' - \frac{\mu}{\kappa} \right)^2\right\}   \, dy' \right)
\end{align*}
After some rearrangement and variable renaming this becomes the \emph{OUP/LIF Mean First Passage Time} (OUP-MFPT) formula,

\bigskip
{\bf \ \   \quad \quad \quad \ \ Siegert formula for the OUP-MFPT} 
\begin{align}
\boxed{ \quad m_1(\theta, y) = \frac{\sqrt{\pi}}{\kappa}\int\limits_{\frac{y - \frac{\mu}{\kappa}}{\sigma/\sqrt{\kappa}}}^{\frac{\theta- \frac{\mu}{\kappa}}{\sigma/\sqrt{\kappa}}} e^{\frac{\ \, z^2}{2}}\left( 1 + \text{\bf erf} (z) \right) dz  = 
\frac{\sqrt{\pi}}{\kappa}\int\limits_{\frac{\kappa y - \mu}{\sigma \sqrt{\kappa \, }}}^{{\frac{\kappa \theta- \mu}{\sigma \sqrt{\kappa \, }}}} e^{\frac{\ \, z^2}{2}}\left( 1 + \text{\bf erf} (z) \right) dz \quad } \label{eq:m1siegert}
\end{align}
where {\bf erf} denotes the error function. 
This result is well known in the literature.  Reference  \cite{lSacerdote:2013a} refers to it as the \emph{Siegert formula}, despite its specialization to the case of the first moment, because it is readily derived from the general Siegert formula \eqref{eq:siegert2} as we have done here.

\medskip
Via use of the general Siegert formula \eqref{eq:siegert2} one can now proceed to derive the second moment of the first passage time for the OUP, and hence the variance of the passage time from knowledge of both the first and second moments. The form of the solutions (though not the details of the computations) can be found in \cite{muThomas:1975a}, though represented in term of the normal integral (which is closely related to the error function).

\section{Mean Firing Rate for the OUP/LIF Neuron}

Exploiting the fact that under the assumptions described in Section 1 the LIF subthreshold dynamics describe an Ornstein-Uhlenbeck process (OUP), we have been able to derive the mean first passage time (MFPT) for the neuron to hit the firing threshold after a reset at time $t = 0$. Specifically, MFPT as a function of the threshold value $\theta$ and the membrane voltage reset value $v_0$ is given by Eq.~\eqref{eq:m1siegert},
\[  \widehat{T} = \widehat{T}_\theta(x_0) = \text{E} \{ T_\theta(x_0) \}  = m_1(\theta,v_0) . \]
As discussed above, once the MFPT is at hand, the {mean time-to-fire} (MTF) of the OUP/LIF neuron,
\[ \widehat{T}_f = \text{E}(T_f) = T_r + \text{E}(T) = T_r + \widehat{T}, \]
 is used to form the  {average firing rate} (AFR) of the neuron,
\[ \rho = \widehat{T}_f^{-1} = \frac{1}{T_r + \widehat{T}} . \]

Recall from Eq.~\eqref{eq:alldefined} that of all the parameters that define the OUP/LIF neuron, only 
\begin{align}
 \mu = \mu(I_0) = \frac{1}{\tau} V_r +  \frac{1}{C}\widehat{I}_{\rm syn}  + I_0 \label{eq:mui0}
\end{align}
is generally under our immediate control.  It is evident that the first two terms on the right-hand-side of this equation can have a confounding effect that must be compensated for. Given the relationship between $\mu(I_0)$ and the injected current \eqref{eq:mui0}, and recalling that,
\[ \kappa = \frac{1}{\tau} = \frac{1}{RC} , \]
we obtain the \emph{Siegert formula} which shows how $I_0$ affects \emph{the mean firing rate for the OUP/LIF Neuron} (LIF-MFR) in terms of the parameters defining the LIF-MFR neuron,

\bigskip
{\bf \qquad \qquad \qquad  \quad \,  \, \  \ Siegert formula for the LIF-MFR} 
\begin{align}
\boxed{ 
\begin{aligned}
\rho(I_0)  & = \left(T_r +  \tau \, \sqrt{\pi }  \int\limits_{\frac{ y - \tau \mu(I_0)}{\sqrt{\tau \, }\sigma}}^{{\frac{ \theta- \tau \mu(I_0)}{ \sqrt{\tau \, } \sigma}}} e^{\frac{\ \, z^2}{2}}\left( 1 + \text{\bf erf} (z) \right) dz \right)^{-1} \\
\mu(I_0) & = \frac{1}{\tau} V_r +  \frac{1}{C}\widehat{I}_{\rm syn}  + I_0 , \quad \tau = RC \label{eq:lifmfr}
\end{aligned}} 
\end{align}

\medskip
As mentioned in the first section of this note, the OUP-LIF is a tremendously simplified model of neuronal dynamics.  There is much research activity involved with relaxing the model assumptions.  As one can imagine, determining the mean first passage time for more complex is a highly nontrivial endeavor.  A low-order correction to the LIF-MFR \eqref{eq:lifmfr} to partially account for synaptic capacitive effects is given in \cite{nBrunel:1998a}.  Further relevant modeling issues are discussed in the survey paper \cite{aRenart:2003a}.


\bigskip
\bigskip

\setcounter{equation}{0}  
\renewcommand{\theequation}{A.\arabic{equation}} 

\noindent{\large \bf Appendix -- Stationary Distribution of the Unrestricted FPE}

\bigskip
The Markovian probabilistic behavior of the homogeneous stochastic differential equation \eqref{eq:sde1} is described by the transition probability density function, $p( x, t | x_0 , t_0)$, which is a solution to the (forward) Fokker-Planck Equation \eqref{eq:FPE2}
subject to the initial condition \eqref{eq:deltaic} and the unrestricted boundary conditions \eqref{eq:unrestrictedbc}.  Under reasonable conditions on the SDE \eqref{eq:sde1}, in the limit as $t \to \infty$ the pdf $p( x, t | x_0 , t_0)$ converges to a steady-state density, $\mathsf{W}(x)$, which is independent of the time, $t$, and of the initial information $(x_0,t_0)$ \cite{cwGardiner:2009a,vanKampen:2007a},
\begin{align}
\mathsf{W}(x) = \lim_{t \to \infty}  p( x, t | x_0 , t_0) . 
\end{align}
Since in the limit the steady-state density $\mathsf{W}(x)$ is independent of the initial condition \eqref{eq:deltaic}, and $\frac{\partial}{\partial t} \mathsf{W}(x) = 0$, it must solve the steady-state (forward) Fokker-Planck equation,
\begin{align}
  - \frac{\partial}{\partial x}\alpha(x)  \mathsf{W}(x) +  \frac{1}{2} \frac{\partial^2}{\partial x^2} \beta(x) \mathsf{W}(x) = 0  ,  \label{eq:Aeqss}
\end{align}
subject to the boundary conditions \eqref{eq:unrestrictedbc} and the pdf normalization condition,
\begin{align}
\int \mathsf{W}(x) dx = 1 . 
\end{align}

\medskip
Note that we can rewrite Equation \eqref{eq:Aeqss} as
\[ \frac{\partial}{\partial x}\left[ \frac{\partial }{\partial x }  \Big( \beta(x) \mathsf{W}(x)\Big) - 2  \alpha(x)  \mathsf{W}(x) \right] = 0 , \]
which yields,
\[ \frac{\partial }{\partial x }  \Big( \beta(x) \mathsf{W}(x)\Big) - 2  \alpha(x)  \mathsf{W}(x)  = 0, \]
assuming that,\footnote{Note that these are stronger conditions than merely requiring that $\mathsf{W}(-\infty) = 0$.  They correspond to conditions being placed on $\alpha(x)$ and $\beta(x)$ (i.e., on the SDE \eqref{eq:sde1}).} 
\[ \left( \frac{\partial }{\partial x }  \Big( \beta(x) \mathsf{W}(x)\Big)\right) \bigg|_{x = - \infty}  =   \Big( \alpha(x)  \mathsf{W}(x) \Big) \bigg|_{x = - \infty}  = 0 . \]
An additional integration yields,
\begin{align}
\mathsf{W}(x) = \frac{C}{\beta(x)} \exp\left\{  2 \int_{-  \infty}^x \frac{\alpha(x')}{\beta(x')}  dx' \right\} \, , \label{eq:W2}
\end{align}
assuming that,
\[  \Big( \beta(x) \mathsf{W}(x)\Big)\bigg|_{x = - \infty} = 0 ,\]
where $C$ is a constant of integration.  

\medskip
Note the equivalence of equations \eqref{eq:W1} and \eqref{eq:W2}.  If we chose the constant $C$  to ensure that $\mathsf{W}(x)$ is a properly normalized probability density function, we obtain the following pdf as a steady-state solution to the Fokker-Planck equation,
\begin{align}
\mathsf{W}(x) = \frac{\beta(x)^{-1}\exp\left\{  2 \int\limits_{-  \infty}^x \frac{\alpha(x')}{\beta(x')}  dx' \right\} }{\int\limits_{-\infty}^\infty \beta(x'')^{-1} \exp\left\{  2 \int\limits_{-  \infty}^{x''} \frac{\alpha(x')}{\beta(x')}  dx' \right\} dx''} \,  . \label{eq:ssoupdensity}
\end{align}

\medskip
For the Ornstein-Uhlenbeck process (OUP)  \eqref{eq:sdemodel}--\eqref{eq:ouprocess} we have,
\[ \alpha(x) = \alpha_0 - \alpha_1 x  = \mu - \kappa x \quad \text{and} \quad \beta(x) = \beta_0 = \sigma^2 \, . \]
This yields,
\begin{align*}
 \mathsf{W}(x) 
 & = \lim_{\ell \to - \infty} \, \frac{\exp\left\{ \frac{2}{\sigma^2} \int\limits_{  \ell}^x \left(\mu - \kappa \, x' \right) dx' \right\} }{\int\limits_{- \infty}^\infty \exp\left\{  \frac{2}{\sigma^2}  \int\limits_{\ell}^{x''} \left(\mu - \kappa \, x' \right)  dx' \right\} dx''}  \\
& = \lim_{\ell \to - \infty} \, \frac{\exp\left\{ \frac{2}{\sigma^2} \left( \int\limits_{0}^x \left(\mu - \kappa \, x' \right) dx'  + \int\limits_{  \ell}^0 \left(\mu - \kappa \, x' \right) dx'\right) \right\} }{\int\limits_{- \infty}^\infty \exp\left\{  \frac{2}{\sigma^2} \left( \int\limits_{0}^{x''} \left(\mu - \kappa \, x' \right)  dx'  + \int\limits_{  \ell}^0 \left(\mu - \kappa \, x' \right) dx'\right) \right\} dx''} \\
& = \frac{\exp\left\{ \frac{2}{\sigma^2}  \int\limits_{0}^x \left(\mu - \kappa \, x' \right) dx'   \right\} }{\int\limits_{- \infty}^\infty \exp\left\{  \frac{2}{\sigma^2}  \int\limits_{0}^{x''} \left(\mu - \kappa \, x' \right)  dx'   \right\} dx''} \\
& = \frac{\exp\left\{ \frac{2}{\sigma^2}  \left(\mu \, x -  \frac{1}{2} \kappa \, x^2\right)  \right\} }{\int\limits_{- \infty}^\infty \exp\left\{  \frac{2}{\sigma^2}   \left(\mu\, x''-  \frac{1}{2} \kappa \, x''^2 \right)   \right\} dx''} \, . 
 \end{align*}
Now note that,
\[ 2 \left(\mu \, x - \frac{1}{2} \kappa \, x^2 \right) = - \kappa \left( x^2 - 2 \frac{\mu}{\kappa} \, x  +\frac{\mu^2}{\kappa^2} - \frac{\mu^2}{\kappa^2}\right)  = - \kappa \left(x - \frac{\mu}{\kappa} \right)^2 +  \frac{\mu^2}{\kappa} . \]
Thus,
\[ \mathsf{W}(x) = \frac{\exp\left\{ - \frac{\kappa}{\sigma^2}  \left(x -  \frac{\mu}{\kappa} \right)^2  \right\} }{\int\limits_{- \infty}^\infty \exp\left\{ -  \frac{\kappa}{\sigma^2}   \left( x''-  \frac{\mu}{\kappa}  \right)^2   \right\} dx''} \, , \]
or,
\begin{align}
\mathsf{W}(x) =  \sqrt{\frac{\kappa}{\pi \sigma^2}} \, \exp\left\{ - \frac{\kappa}{\sigma^2}  \left(x -  \frac{\mu}{\kappa} \right)^2  \right\} ,
\end{align} 
which is a Gaussian pdf with mean $\frac{\mu}{\kappa}$ and variance $\frac{\sigma^2}{2\, \kappa}$.  This is a standard result.\footnote{E.g., see \cite{hcTuckwell:1988v2}.}


\bibliographystyle{plain}
\bibliography{KKD.bib} 


\end{document}